\documentclass[conference]{IEEEtran}
\usepackage[top=0.75in, bottom=1.0in, left=0.625in, right=0.625in]{geometry}
\IEEEoverridecommandlockouts
\usepackage{cite}
\usepackage{amsmath,amssymb,amsfonts}
\usepackage{algorithmic}
\usepackage{graphicx}
\usepackage{float}
\usepackage{placeins}
\usepackage{textcomp}
\usepackage{xcolor}
\usepackage[cmintegrals]{newtxmath}
\usepackage{bm}
\usepackage{acro}
\usepackage{enumitem}
\usepackage{dblfloatfix}
\usepackage{ragged2e} 

\usepackage{gensymb}
\usepackage{ragged2e}

\usepackage{tabularx}
\usepackage{makecell}

\usepackage{array}
\usepackage{multirow}

\usepackage{subfig}
\usepackage{subcaption}

\usepackage{adjustbox} 
\usepackage{booktabs}
\usepackage{colortbl} 
\usepackage{xcolor} 

\usepackage[font=footnotesize]{caption}

\usepackage{fancyhdr}
\fancypagestyle{firststyle}{
    \fancyhf{}
    \fancyhead[L]{I. Jariwala, X. Wang, B. Meier, G. Qian, D. Shakya, M. Ying, H. Nikbakht, D. Abraham, and T. S. Rappaport, ``NYUSIM: A Roadmap to AI-Enabled Statistical Channel Modeling and Simulation,'' to appear in \textit{IEEE International Conference on Communications (ICC)}, Glasgow, UK, Jun. 2026, pp. 1--6.}
}

\newcolumntype{?}{!{\vrule width 1pt}}

\definecolor{lightgray}{gray}{0.95}

\usepackage{titlesec}
\titlespacing*{\section}{0pt}{0.8ex plus .2ex minus .2ex}{0.5ex plus .2ex}

\def\BibTeX{{\rm B\kern-.05em{\sc i\kern-.025em b}\kern-.08em
    T\kern-.1667em\lower.7ex\hbox{E}\kern-.125emX}}

\DeclareAcronym{ai}{
  short = AI,
  long  = Artificial Intelligence
}

\DeclareAcronym{ml}{
  short = ML,
  long  = Machine Learning
}

\DeclareAcronym{sscm}{
  short = SSCM,
  long  = Statistical Spatial Channel Model
}

\DeclareAcronym{tcsl}{
  short = TCSL,
  long  = Time-Cluster Spatial-Lobe
}

\DeclareAcronym{mmwave}{
  short = mmWave,
  long  = Millimeter Wave
}

\DeclareAcronym{subthz}{
  short = sub-THz,
  long  = Sub-Terahertz
}

\DeclareAcronym{fr3}{
  short = FR3,
  long  = Frequency Range 3 (7–24 GHz)
}

\DeclareAcronym{fr1c}{
  short = FR1(C),
  long  = Frequency Range 1(C) – Upper Portion of FR1 (~6–8 GHz)
}

\DeclareAcronym{pdp}{
  short = PDP,
  long  = Power Delay Profile
}

\DeclareAcronym{cdf}{
  short = CDF,
  long  = Cumulative Distribution Function
}

\DeclareAcronym{pdf}{
  short = PDF,
  long  = Probability Density Function
}

\DeclareAcronym{rmsds}{
  short = RMS DS,
  long  = Root-Mean-Square Delay Spread
}

\DeclareAcronym{los}{
  short = LOS,
  long  = Line-of-Sight
}

\DeclareAcronym{nlos}{
  short = NLOS,
  long  = Non-Line-of-Sight
}

\DeclareAcronym{umi}{
  short = UMi,
  long  = Urban Microcell
}

\DeclareAcronym{uma}{
  short = UMa,
  long  = Urban Macrocell
}

\DeclareAcronym{rma}{
  short = RMa,
  long  = Rural Macrocell
}

\DeclareAcronym{inf}{
  short = InF,
  long  = Indoor Factory
}

\DeclareAcronym{isac}{
  short = ISAC,
  long  = Integrated Sensing and Communication
}

\DeclareAcronym{ns3}{
  short = ns-3,
  long  = Network Simulator 3
}

\DeclareAcronym{3dant}{
  short = 3Dant,
  long  = Three-Dimensional Antenna Format
}

\DeclareAcronym{hpbw}{
  short = HPBW,
  long  = Half-Power Beamwidth
}

\DeclareAcronym{gan}{
  short = GAN,
  long  = Generative Adversarial Network
}

\DeclareAcronym{vae}{
  short = VAE,
  long  = Variational Autoencoder
}

\DeclareAcronym{dnn}{
  short = DNN,
  long  = Deep Neural Network
}

\DeclareAcronym{ci}{
  short = CI,
  long  = Close-In Reference Distance
}

\DeclareAcronym{threed}{
  short = 3D,
  long  = three-dimensional
}

\DeclareAcronym{ks}{
  short = K-S,
  long  = Kolmogorov–Smirnov
}
\begin{document}

\bstctlcite{BSTcontrol}
     \title{NYUSIM: A Roadmap to AI-Enabled Statistical Channel Modeling and
Simulation}

\author{\IEEEauthorblockN{%
Isha Jariwala\textsuperscript{1},
Xinquan Wang,
Bridget Meier,
Guanyue Qian,\\
Dipankar Shakya,
Mingjun Ying,
Homa Nikbakht\textsuperscript{2},
Daniel Abraham, and
Theodore~S.~Rappaport\textsuperscript{3}}
\IEEEauthorblockA{New York University, Tandon School of Engineering, Brooklyn, NY, USA\\
\{\,\textsuperscript{1}ij2221,
\textsuperscript{2}homa.n,
\textsuperscript{3}tsr\,\}@nyu.edu}}
\maketitle

\thispagestyle{firststyle}
\begin{abstract}
Integrating artificial intelligence (AI) into wireless channel modeling requires large, accurate, and physically consistent datasets derived from real measurements. Such datasets are essential for training and validating models that learn spatio-temporal channel behavior across frequencies and environments. NYUSIM, introduced by NYU WIRELESS in 2016, generates realistic spatio-temporal channel data using extensive outdoor and indoor measurements between 28 and 142 GHz. To improve scalability and support 6G research, we migrated the complete NYUSIM framework from MATLAB to Python, and are incorporating new statistical model generation capabilities from extensive field measurements in the new 6G upper mid-band spectrum at 6.75 GHz (FR1(C)) and 16.95 GHz (FR3) \cite{ojcoms_fr3_overview}. The NYUSIM Python also incorporates a 3D antenna data format, referred to as \emph{Ant3D}, which is a standardized, full-sphere format for defining canonical, commercial, or measured antenna patterns for any statistical or site-specific ray tracing modeling tool. Migration from MATLAB to Python was rigorously validated through Kolmogorov–Smirnov (K–S) tests, moment analysis, and end-to-end testing with unified randomness control, confirming statistical consistency and reproduction of spatio-temporal channel statistics, including spatial consistency with the open-source MATLAB NYUSIM v4.0 implementation. The NYUSIM Python version is designed to integrate with modern AI workflows and enable large-scale parallel data generation, establishing a robust, verified, and extensible foundation for future AI-enabled channel modeling. 
\end{abstract}

\begin{IEEEkeywords}
NYUSIM, FR3, FR1(C), Artificial Intelligence (AI), Machine Learning (ML), 6G, 3D antenna modeling, statistical spatial channel modeling
\end{IEEEkeywords}

 \section{Introduction and Background}
 Reliable channel models form the foundation for the design, test, and standardization of wireless communication systems. Over the past decade, the demand for accurate, measurement-based channel modeling has increased as mobile networks evolve toward higher frequencies, wider bandwidths, and greater spatial complexity in antenna radiation patterns. The NYUSIM channel simulator, introduced by NYU WIRELESS in 2016 \cite{poddar2024tutorial, RappaportVTC2017Compare3GPPNYUSIM, SunICC2017NYUSIM}, has become a global reference platform for studying millimeter-wave (mmWave), sub-Terahertz (sub-THz), and now upper mid-band (FR3) propagation environments. Built on more than a decade of outdoor and indoor measurement campaigns, spanning 28 to 142 GHz, NYUSIM implements the statistical spatial channel model (SSCM) using the time cluster spatial lobe (TCSL) framework \cite{Samimi2015,Rappaport2015Com}. NYUSIM captures the physical behavior of multipath propagation, reproducing large-scale path loss (PL), multipath delay, angular spreads, and spatio-temporal power distributions observed in real measurements. Since its public release, NYUSIM has become one of the most widely used open-source propagation tools, downloaded over 100,000 times, cited in more than 3,300 publications (as of 2024), and incorporated into network-level simulators such as ns-3 \cite{PoddarICC2023,PoddarWS2023}.


The evolution of NYUSIM, beyond version 4.0 as presented here, describes three major advancements: (1) extension to FR3 frequencies, (2) inclusion of realistic three-dimensional (3D) antenna pattern modeling, and (3) complete conversion of the MATLAB code base into a modular and open Python framework with clearly separated functional components. The FR3 band (7–24 GHz) represents the “upper mid-band” of the spectrum expected to support global 6G deployment \cite{ojcoms_fr3_overview}.
Recent actions by the ITU, NTIA, FCC, and WRC-23 have highlighted specific FR3 sub-bands (e.g., $7.125$-$8.4$ GHz, $14.8$-$15.35$ GHz) as strong candidates for future mobile allocations \cite{ojcoms_fr3_overview}. Emerging use cases in 6G and integrated sensing and communication (ISAC) applications further increase the need for accurate, measurement-based FR3 channel models\cite{bazzi2025uppermidband}.
Previous upper mid-band work \cite{oyie2018_14_22GHz, kim2014_11GHz} reported indoor line-of-sight (LOS) and non-line-of-sight (NLOS) path-loss exponents and delay spreads at selected frequencies (e.g., $6–14$ GHz), but datasets were limited in bandwidth, environment, or angular coverage. In contrast, 
NYU WIRELESS conducted the world’s first comprehensive measurement campaigns pairing $6.75$ GHz (FR1(C)), and $16.95$ GHz (FR3)~\cite{ojcoms_fr3_overview,globec_indoors,icc_inf,globec_penetration}.

Accurate statistical channel modeling in FR1(C) and FR3 requires realistic embedded antenna patterns~\cite{Sulyman2014,3gpp38901}. Here, we develop a 3D antenna data format, referred to as \emph{Ant3D}. \emph{Ant3D} is a standardized, full-sphere format for defining canonical, commercial, or measured antenna patterns for any statistical or site-specific (e.g., ray-tracing) modeling tools. Each antenna pattern is represented as a gain matrix over azimuth and elevation with optional frequency components, allowing directional and polarization effects to be incorporated into simulated spatial channel impulse responses. \emph{Ant3D} extends NYUSIM beyond the canonical uniform linear array (ULA), 
enabling channel simulations with realistic array geometries, and sidelobe structures~\cite{NIST2019,Xing140GHz2021,xing2021outdoor}.


NYUSIM 4.0 is ported from MATLAB to Python to provide a reliable, transparent AI-native channel simulator derived from gold-standard measurement data, and it can learn as new propagation data becomes available.
The migration to NYUSIM Python provides a modular, scalable
software architecture to support AI workflows for generative and discriminative channel models, faithfully reproduces real-world channel spatio-temporal sample functions, while allowing future measurements to be seamlessly integrated and learned within this platform~\cite{Rappaport2017TAP,COST2100,QuaDRiGa}.

\section{Motivation to Use AI in Simulation} 
\par

Recent advances in generative and discriminative AI have motivated a paradigm shift from purely statistical channel modeling to data-driven wireless channel synthesis. 
AI-driven modeling uses measured datasets to learn complex, and nonlinear mappings between environmental variables (frequency, scenario, geometry) and channel characteristics (path loss, delay spread, angular spread, cluster dynamics) \cite{Wu2024CDDM_TWC, Dong2019DeepCNN_mmWaveMIMO}.
Generative AI models such as diffusion models, variational autoencoders (VAEs), and generative adversarial networks (GANs) are capable of learning the joint probability distribution of channel parameters \cite{Wu2024CDDM_TWC}. Generative AI allows for realistic synthesis of unseen propagation environments by sampling from measurement-based statistical distributions that capture the spatio-temporal characteristics of real channels. Discriminative models, such as deep neural networks and random forests, are capable of predicting large-scale and small-scale parameters or beamforming statistics from input conditions such as frequency, antenna geometry, and mobility \cite{Dong2019DeepCNN_mmWaveMIMO}.  The approaches in \cite{Wu2024CDDM_TWC, Dong2019DeepCNN_mmWaveMIMO} are increasingly being investigated for 6G digital twins, map-aware simulations, and AI-native PHY/MAC co-design, where analytical interpolation (e.g., linear or parabolic) fails to capture frequency-dependent nonlinearity and spatial correlation \cite{10437154}.

Classical channel simulators, especially closed or proprietary ones, are not easily integrated with AI models for three reasons: 
\begin{itemize}
\item Restricted data access: Measurement datasets and parameter generation are often embedded in compiled code, limiting large-scale data export for AI training.
\item Limited scalability: The software architectures of traditional channel simulators are not scalable, as they are not designed for parallel generation or cloud execution, which makes it difficult to produce millions of labeled and unlabeled samples required by modern AI models.
\item Limited interoperability: Closed systems do not integrate cleanly with  AI frameworks such as PyTorch and TensorFlow, thus preventing direct embedding of learned models into the channel-simulation workflow that generates channel coefficients and impulse responses.
\end{itemize}

The aforementioned shortcomings limit progress in self-evolving modeling, as accurate AI-driven channel simulators require massive datasets across frequency and spatio-temporal dimensions.

A Python implementation of NYUSIM aligns naturally with modern AI and systems workflows, enabling capabilities that are difficult to realize in MATLAB. Python interfaces directly with mainstream ML ecosystems, and NumPy arrays convert to PyTorch/JAX/TensorFlow in one line. Hence,  generative models (diffusion, GANs, VAEs) and discriminative models (for path loss, spreads, blockage, beams) can be trained without intermediate conversions or custom wrappers. Python toolchains also support large-scale, automated data generation and streaming via multiprocessing or cluster frameworks (e.g., Ray/Dask), making supervised and self-supervised training on millions of channel realizations practical. Through Python APIs, NYUSIM integrates with the network simulator (ns-3) for closed-loop experiments in which learning agents observe PHY/MAC states and act within the same ns-3 process\cite{Abadi2016TensorFlow}.

Consider how NYUSIM Python supports the AI workflows described above through a measurement-driven learning pipeline. After collecting field measurements at a new frequency or environment, data is imported directly into the NYUSIM Python framework. An AI model then trains on measurements to automatically learn the underlying statistical patterns, such as how signal power decays with distance, how multipath clusters arrive in time, or how energy spreads across angles. Because the AI learns patterns from the data itself, it can capture propagation behaviors that are difficult to derive manually through traditional curve fitting. The learned model is then plugged into NYUSIM through its modular Python interface, updating the relevant simulation modules. 

 \section{NYUSIM: from MATLAB to Python} \label{sec:conversion}
 The migration of NYUSIM from MATLAB to Python improves accessibility, scalability, and reproducibility in propagation research, as Python enables integration with AI and ML frameworks for data-driven channel modeling. The first four MATLAB versions of NYUSIM established the foundation for the SSCM and have long been used to study mmWave and sub-THz channels~\cite{Samimi2015,Rappaport2015Com,Rappaport2017TAP}. However, extending NYUSIM to new frequency bands or channel conditions required manual modification of the open-source code, and large parameter sweeps or parallel runs were constrained by MATLAB’s proprietary execution model. Integration with external network simulators such as ns-3 also involved custom interfaces and translation layers~\cite{PoddarICC2023,PoddarWS2023}. The new Python framework retains all physical equations, statistical behavior, and input–output structures of the MATLAB version, while introducing a modular architecture that simplifies model expansion, parallel execution, and AI integration~\cite{poddar2024tutorial,letaief2019roadmap}.

 \subsection{NYUSIM Python Framework}
 To ensure consistency with the MATLAB version, we reorganized the channel-generation functions of NYUSIM Python into \emph{deterministic} and \emph{stochastic} classes, a standard split in wireless propagation modeling~\cite{3gpp38901,COST2100,QuaDRiGa}. Deterministic classes (e.g., mean free-space path loss, atmospheric attenuation, frequency scaling) encode fixed physics and are verified by comparing function outputs between the MATLAB and Python versions of NYUSIM. Stochastic classes (e.g., shadow fading, power delay profiles, angular spreads) capture randomness and are validated via Monte Carlo (MC) ensembles by comparing distributions and moments, using seed-controlled draws~\cite{Rappaport2017TAP,Xing140GHz2021,NIST2019,PoddarICC2023,Tranter2004Simulation}. The class separation isolates each physical process for targeted testing and extension, while preserving internal consistency across environments and frequencies.

During the code conversion, each MATLAB function was translated into a Python function with identical input–output structures and parameter definitions. Constants, unit conventions, and random number seeds were preserved to maintain physical accuracy and reproducibility~\cite{Rappaport2015Com}. The matrix operations and vectorized calculations in MATLAB were mapped to equivalent NumPy and SciPy in Python to achieve similar computational behavior while improving expandability. Conditional logic used in MATLAB scripts was refactored into structured Python dictionaries that describe the channel state, such as environment type, transmitter–receiver geometry, and frequency configurations~\cite{poddar2024tutorial}.

The migration of NYUSIM from MATLAB to Python shows that channel simulators can evolve while preserving physical rigor and measurement fidelity. The Python framework underscores a key lesson for the propagation community: accurate modeling relies not only on high-quality measurements but also on transparent and reproducible software architectures that support ongoing validation~\cite{Rappaport2017TAP,poddar2024tutorial} and enable learning from large-scale measurement data through AI and ML methods to refine and extend propagation models. The principles of transparency, reproducibility, and measurement-driven modeling may be applied to upgrade or develop future simulators that enable AI-assisted learning in propagation research~\cite{letaief2019roadmap}.

    \subsection{Verification and Statistical Validation of NYUSIM Python}
We describe frameworks for verification and statistical validation testing of NYUSIM Python.  
\subsubsection{Function-to-Function Testing}
We developed a function-to-function testing framework to confirm that each deterministic and stochastic process in the Python version produces statistically equivalent outputs to its MATLAB counterpart. Deterministic functions were compared element-wise using absolute and relative error thresholds to ensure numerical equivalence between MATLAB and Python outputs. Stochastic functions were statistically validated in more than 10,000 MC realizations using the K–S test and moment-based analysis. The K–S test showed that the empirical distributions of delay spread, angular spread, and shadow fading, were statistically indistinguishable between MATLAB and Python, while the means and variances of NYUSIM Python agreed within $1\%$. Tests were performed across all standardized NYUSIM environments: urban microcell (UMi), urban macrocell (UMa), rural macrocell (RMa), indoor factory (InF), and indoor hotspot (InH), and over frequencies from 28 to 142 GHz under both LOS and NLOS conditions. The function-to-function verification confirms that, for each operating frequency and propagation scenario defined in the MATLAB version, the Python implementation reproduces both deterministic and stochastic behaviors with extremely high statistical fidelity.
\subsubsection{End-to-End Testing of NYUSIM Python Operation}

In the end-to-end testing, MATLAB serves as the baseline, and the Python code is required to mirror it. End-to-end testing aims to verify that the full workflow from inputs to final outputs, such as the number of multipaths, path delays, powers, and RMS delay and angular spreads, matches in overall statistics. The testing is challenging because small numerical and stochastic output differences of each function can accumulate. The main difficulty is random number generation. MATLAB uses MT19937, which is a deterministic pseudo-random generator with period $2^{19937}-1$. While NumPy uses PCG64, a permuted congruential generator (PCGs) \cite{oneill:pcg2014} and a different float-mapping path. As a result, feeding identical seeds to the MATLAB and Python versions of NYUSIM leads to numerically different sequence outputs. 

To address the sequence mismatch, we implemented the same MT19937 in both MATLAB and Python versions of NYUSIM, adopted an identical 53-bit uniform mapping, and routed all random draws (uniform, normal, lognormal, exponential, Poisson, gamma) through functions that mimic calls. We also matched formulas, constants, array shapes, and edge-case rules and accounted for MATLAB 1-based, column-oriented style versus NumPy 0-based indexing. Finally, we executed MATLAB and Python versions on the same element-wise inputs and compared outputs. Test results confirmed that the developed Python implementation reproduces the statistics of MATLAB. In addition, we performed an end-to-end MC test with 10,000 iterations using random seeds. The MC results also confirmed the consistency between the two versions using the K–S test and moment-based analysis. Note that in single-run execution, the Python and MATLAB implementations have comparable runtime and memory usage. The scalability advantage of Python allows parallel simulation across many machines using free tools such as multiprocessing, Ray, or Dask,  without licensing costs, which is important when generating the large datasets needed for AI training.

\section{Implementing 3D Antenna Patterns in NYUSIM} 
 \subsection{Ant3D Format}\label{sec:ant3d}
NYUSIM Python incorporates a unified 3D antenna data format, referred to as \emph{Ant3D}, to represent customizable, commercial, or canonical real antenna radiation behavior in spatial channel simulations. The \emph{Ant3D}
format provides a standardized description of both transmitting and receiving antennas, enabling the use of measured, simulated, or vendor-supplied radiation patterns in a consistent structure~\cite{Rappaport2017TAP,Samimi2015,QuaDRiGa,NIST2019}. 
Each storable antenna pattern is defined on a spherical grid of azimuth ($\phi$) and elevation ($\theta$) angles, covering [$0^{\circ}$, $360^{\circ}$] and [$-90^{\circ}$, $90^{\circ}$], respectively. Azimuth $\phi$ is defined counterclockwise in the horizontal plane, and elevation $\theta$ is measured from the horizontal plane with positive elevation. AOD/AOA map to azimuth, while ZOD/ZOA are converted to elevation via $\theta = 90^\circ - \text{ZOD/ZOA}$. Antenna patterns are applied in simulation as either gain-only patterns or using spherical field components. The format stores antenna gain values in dBi within a matrix $G(\theta,\phi)$. Data structure accommodates isotropic, horn, dipole, and phased-array antennas, as well as imported 3D patterns from commercial antenna specifications, electromagnetic solvers, or measurement campaigns~\cite{Sulyman2014,Xing140GHz2021,xing2021outdoor}. All data in the \emph{Ant3D} antenna pattern format are normalized to the maximum gain of the antenna to ensure consistency across frequencies and scenarios.

The \emph{Ant3D} structure is implemented in Python as a data block with clearly defined \textit{data fields} (i.e., labeled components of the dataset), including frequency, angular grids, gain matrix, polarization, and normalization reference. Each record stores both the normalized gain matrix $G(\theta,\phi)$ and the corresponding peak gain $G_{\text{max}}$, ensuring that directional gain information is preserved for every frequency or scenario. Data blocks also describe the orientation, placement configuration, operating frequencies and 3D antenna pattern data source of the antenna.

During simulation, NYUSIM Python interpolates gain values for the randomly generated arrival and departure angles and applies orientation alignment at both the transmitter and receiver. The resulting direction-dependent gain modifies the power of each multipath component, capturing the effects of beam directivity, sidelobes, and polarization mismatch on the channel impulse response. By standardizing antenna representation, the \emph{Ant3D} format enables reproducible evaluation of antenna-dependent propagation in FR3, mmWave, and sub-THz frequencies~\cite{3gpp38901,Rappaport2015Com}.






\vspace{-10pt}

\subsection{3D Reconstruction of Antenna Patterns}

To allow NYUSIM users to employ practical commercial antenna patterns sold throughout the industry and across emerging FR3 and higher frequencies, a new antenna pattern (AntPat) module is used in NYUSIM Python. The AntPat module provides a unified workflow to import vendor antenna patterns from \emph{Ant3D} formatted data (as described in Sec. \ref{sec:ant3d}) as well as to alternatively generate 3D patterns that comply with 3GPP or other standards. The AntPat module provides a single representation on the unit sphere that can be visualized, stored, and consumed by the NYUSIM v4.0 channel simulation framework as described in \cite{PoddarICC2023,PoddarWS2023,poddar2024tutorial,nyusimv4manual, SamimiTMTT2016_3DStatModel,RappaportVTC2017Compare3GPPNYUSIM,SunTVT2018PropagationModels,SunICC2017NYUSIM}. Specifically, NYUSIM users can select a specific 3GPP or commercial antenna at both TX and RX when simulating channels with the Python version.

The NYUSIM Python AntPat module includes several commercial antenna patterns from different vendors, where additional vendors and antenna patterns are easily implemented using the AntPat module. However, vendor documentation commonly provides principal cuts only: a vertical cut (V-cut) $G_{\mathrm{V}}(\theta)$ at $\phi=0^\circ$ and a horizontal cut (H-cut) $G_{\mathrm{H}}(\phi)$ at $\theta=0^\circ$, together with catalog peak gain and half-power beamwidth (HPBW) values. The AntPat module reconstructs a full 3D radiation pattern from these parameter inputs through an AntPat reconstruction pipeline that remains faithful to the provided measurements by the vendor.

To upload a vendor's antenna pattern in NYUSIM Python, the user simply enters the V-cut and H-cut data in the required tabular format with angle (degrees) versus gain (dBi) for each cut, and then imports the table into the NYUSIM Python AntPat module.
The AntPat reconstruction procedure first reads the two plane cuts (V-cut and H-cut), removes duplicate angles, and sorts the data. Both cuts are then mapped to a uniform spherical grid by the NYUSIM Python AntPat module. The 3D gain is synthesized from the orthogonal cuts using the multiplication method from \cite{ITUM2101}. In the decibel scale,
\begin{equation}
    G(\theta,\phi) \,\text{[dBi]}  \approx G_{\mathrm{H}}(\phi) \text{[dBi]}+ G_{\mathrm{V}}(\theta) \text{[dBi]} - G_{\max} \,\text{[dBi]},
\end{equation}
where $G_{\mathrm{H}}(\phi)$\ and $G_{\mathrm{V}}(\theta)$\ are the measured H-cut gain and V-cut gain in dBi, and $G_{\max}$\,(dBi) is the catalog peak gain reported by the vendor. In linear scale, each linear cut gain divided by $10^{G_{\max}/10}$ yields a weighting factor between zero and one (since each cut gain $\leq G_{\max}$\,(dBi)), and the product of these two weighting factors multiplied by $10^{G_{\max}/10}$ yields the linear 3D gain, which is therefore always non-negative and bounded by the peak gain. In linear scale, $g(\theta,\phi)=10^{G(\theta,\phi)/10}$ is normalized to satisfy
\begin{equation}
\int_{0}^{2\pi}\!\!\int_{0}^{\pi} g(\theta,\phi)\sin\theta\, d\theta d\phi = 4\pi,
\end{equation}
Here, $g(\theta,\phi)$ denotes the absolute linear antenna gain, with peak value $10^{G_{\max}\,\text{[dBi]}/10}$. Finally, the AntPat module maps the gain to field components on the $(\hat{\theta},\hat{\phi})$ basis and exports the grid to the \emph{Ant3D} format with entries $\phi$ (rad), $\theta$ (rad), $E_{\phi}^{\mathrm{re}}, E_{\phi}^{\mathrm{im}}, E_{\theta}^{\mathrm{re}}, E_{\theta}^{\mathrm{im}}$.

The AntPat module also generates standard-compliant patterns directly from specification parameters such as defined in Table 7.3-1 of 3GPP TR 38.901 \cite{3gpp38901}. Let $\theta'' \in [0^\circ,180^\circ]$ denote the zenith angle and $\phi'' \in [-180^\circ,180^\circ]$ denote the azimuth angle. The vertical and horizontal cuts in the decibel scale are
\begin{align}
A_{\mathrm{dB}}^{\mathrm{V}}(\theta'',\phi''\!=\!0^\circ) &= -\min\!\left\{\, 12\!\left(\frac{\theta''-90^\circ}{\theta_{3\mathrm{dB}}}\right)^{\!2}, \mathrm{SLA}_{\mathrm{v}} \right\}, \label{eq:3gppV}\\
A_{\mathrm{dB}}^{\mathrm{H}}(\theta''\!=\!90^\circ,\phi'') &= -\min\!\left\{\, 12\!\left(\frac{\phi''}{\phi_{3\mathrm{dB}}}\right)^{\!2}, A_{\max} \right\}, \label{eq:3gppH}
\end{align}
and thus the 3D pattern is
\begin{equation}
A_{\mathrm{dB}}(\theta'',\phi'') = -\min\!\left\{\, -\big(A_{\mathrm{dB}}^{\mathrm{V}}(\theta'',0^\circ)+A_{\mathrm{dB}}^{\mathrm{H}}(90^\circ,\phi'')\big), A_{\max}\right\}. \label{eq:3gpp3D}
\end{equation}
For the parameters in \eqref{eq:3gpp3D}, we use default settings in \cite{3gpp38901}: $\theta_{3\mathrm{dB}}=65^\circ$, $\phi_{3\mathrm{dB}}=65^\circ$, $\mathrm{SLA}_{\mathrm{v}}=30\,\mathrm{dB}$, $A_{\max}=30\,\mathrm{dB}$, and element peak gain $G_{E,\max}=8\,\mathrm{dBi}$. NYUSIM users can also customize these parameters and export the grid to the \emph{Ant3D} format. 

\begin{figure}[t]
  \centering
  \includegraphics[width=0.8\linewidth]{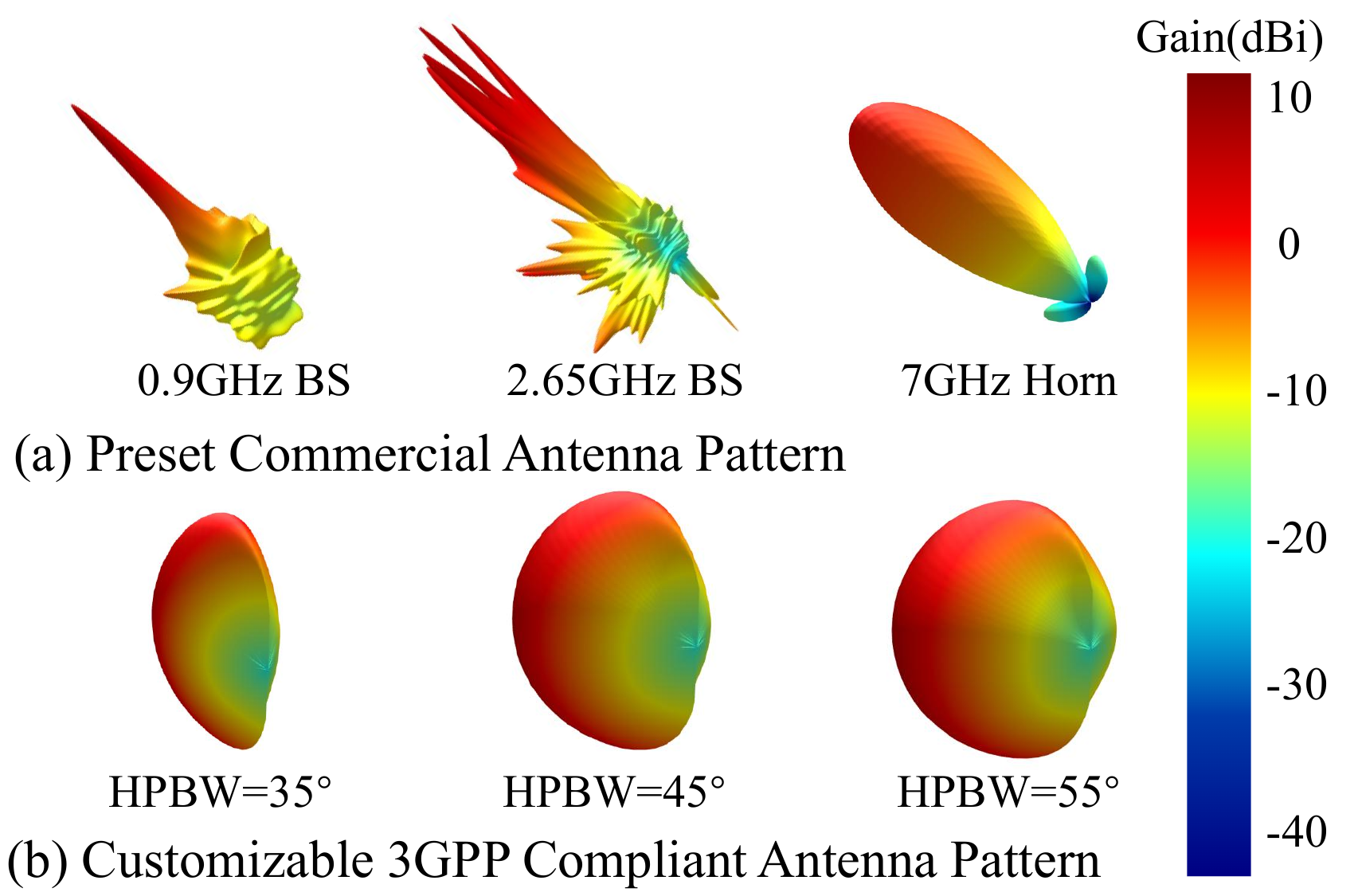}
  \captionsetup{font=footnotesize, name={Fig.}, labelsep=period}
  \caption{Examples of the AntPat module in NYUSIM Python.
  (a) Preset commercial antenna patterns (e.g., JMA iV02OMI136 0.9/2.65 GHz and Pasternack PENWAN-137 antennas).
  (b) Customizable 3GPP-compliant element/array patterns with adjustable parameters.}
  \label{fig:antpat}
  \vspace{-15pt}
\end{figure}
\subsection{Integration of AntPat module with NYUSIM Python}
To integrate antenna patterns into the NYUSIM Python simulation workflow, we implemented a multi-step process as follows.

\subsubsection{Interactive Visualization of Antennas in NYUSIM Python}
To allow quick validation of beamwidth, sidelobe region, and polarization settings before simulation, the AntPat module provides interactive 3D inspection using Python. The interface renders $A(\theta,\phi)$ or $g(\theta,\phi)$ on a sphere or as a surface over $(\theta,\phi)$, with hover inspection, color scales, and camera controls, as shown in Fig. \ref{fig:antpat}. 

\subsubsection{Integration of 3D antenna pattern into Simulation}
We integrate the 3D AntPat module into NYUSIM Python as a post-generation spatial filter applied to the full omnidirectional multipath list.
\begin{itemize}
    \item In each simulation, the NYUSIM Python first generates $L$ omnidirectional spatial-temporal impulse responses with angle-dependent multipath components using same method in NYUSIM v4.0, as described in \cite{PoddarICC2023,PoddarWS2023,poddar2024tutorial,nyusimv4manual,SamimiTMTT2016_3DStatModel,RappaportVTC2017Compare3GPPNYUSIM,SunTVT2018PropagationModels,SunICC2017NYUSIM} and validated in Sec. \ref{sec:conversion}.
    \item Then, NYUSIM Python uses the 3GPP or commercial antenna at TX and RX per user's selection, evaluating each path’s AoD/AoA angles. Specifically, the AntPat module evaluates the chosen 3D patterns at each path’s AoD/AoA to obtain antenna gains $G_{\mathrm t}(\theta_\ell^{\mathrm t},\phi_\ell^{\mathrm t})$ and $G_{\mathrm r}(\theta_\ell^{\mathrm r},\phi_\ell^{\mathrm r})$.
    \item Finally, for each path, the directional path power is computed using the same method in NYUSIM v4.0 \cite{PoddarICC2023,PoddarWS2023,poddar2024tutorial,nyusimv4manual,SamimiTMTT2016_3DStatModel,RappaportVTC2017Compare3GPPNYUSIM,SunTVT2018PropagationModels,SunICC2017NYUSIM}.
\end{itemize}
This integration allows antenna pattern effects on the directional PL to be quantified while maintaining the NYUSIM v4.0 framework \cite{PoddarICC2023,PoddarWS2023,poddar2024tutorial,nyusimv4manual,SamimiTMTT2016_3DStatModel,RappaportVTC2017Compare3GPPNYUSIM,SunTVT2018PropagationModels,SunICC2017NYUSIM}.


 \section{FR3 Modeling in NYUSIM Python}



In 2024, NYU WIRELESS conducted the first comprehensive propagation measurement campaigns in New York City at $6.75$ GHz (FR1(C)) and $16.95$ GHz (FR3) using a $1$ GHz wideband sliding-correlation sounder with dual co-located RF front-ends, precise PTP-synchronized rubidium clocks, and full azimuth/elevation horn-antenna sweeps, capturing tens of thousands of PDPs across LOS/NLOS conditions \cite{globec_indoors,icc_inf,globec_penetration}. 

\begin{figure}[!t]
  \centering
  \includegraphics[width=0.8\linewidth]{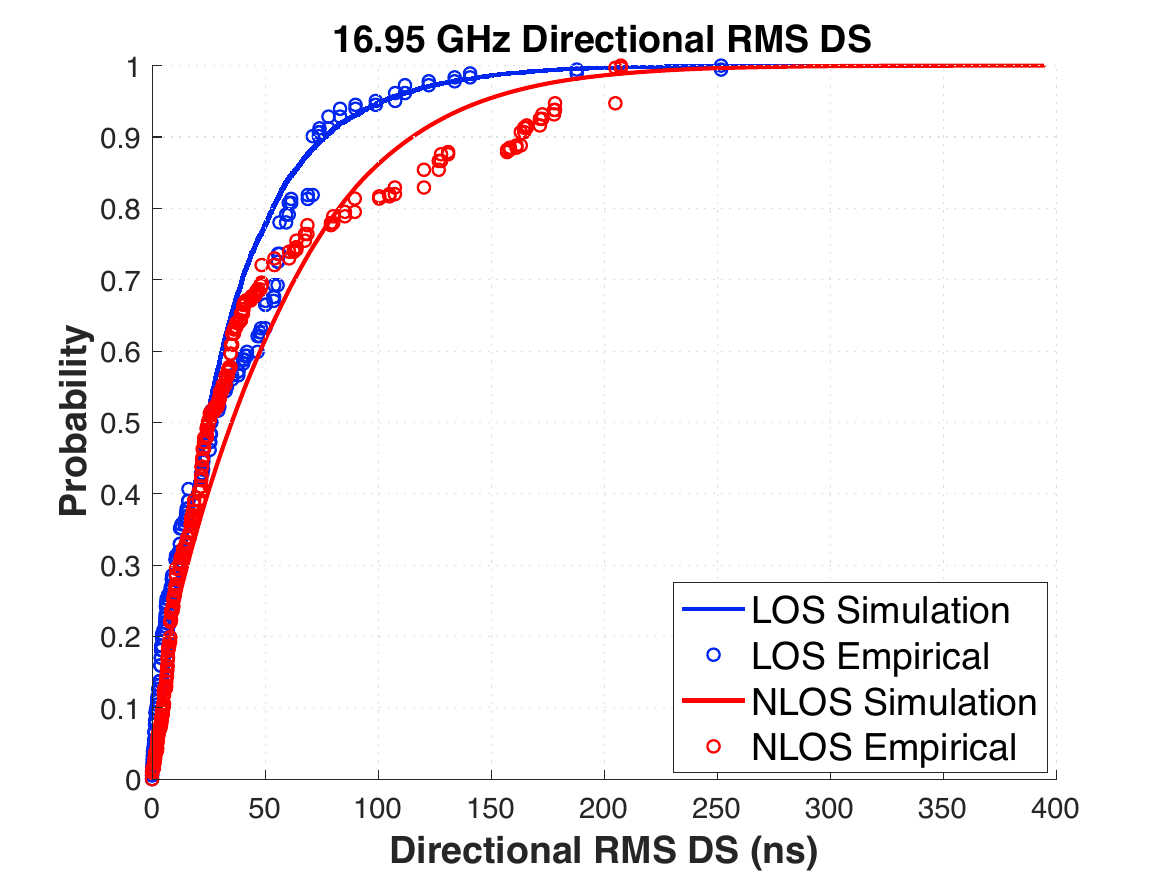}
    \vspace{-3pt}
  \caption{CDF of directional RMS DS for 16.95 GHz. Measurements \cite{globec_indoors,icc_inf,globec_penetration} were conducted using a 1 GHz-bandwidth sliding-correlation channel sounder and directional horn antennas with 20 dBi gain and 15° HPBW.}
  \label{fig:17G}
  \vspace{-0.35cm}
\end{figure}

\begin{figure}[!t]
  \centering
  \includegraphics[width=0.8\linewidth]{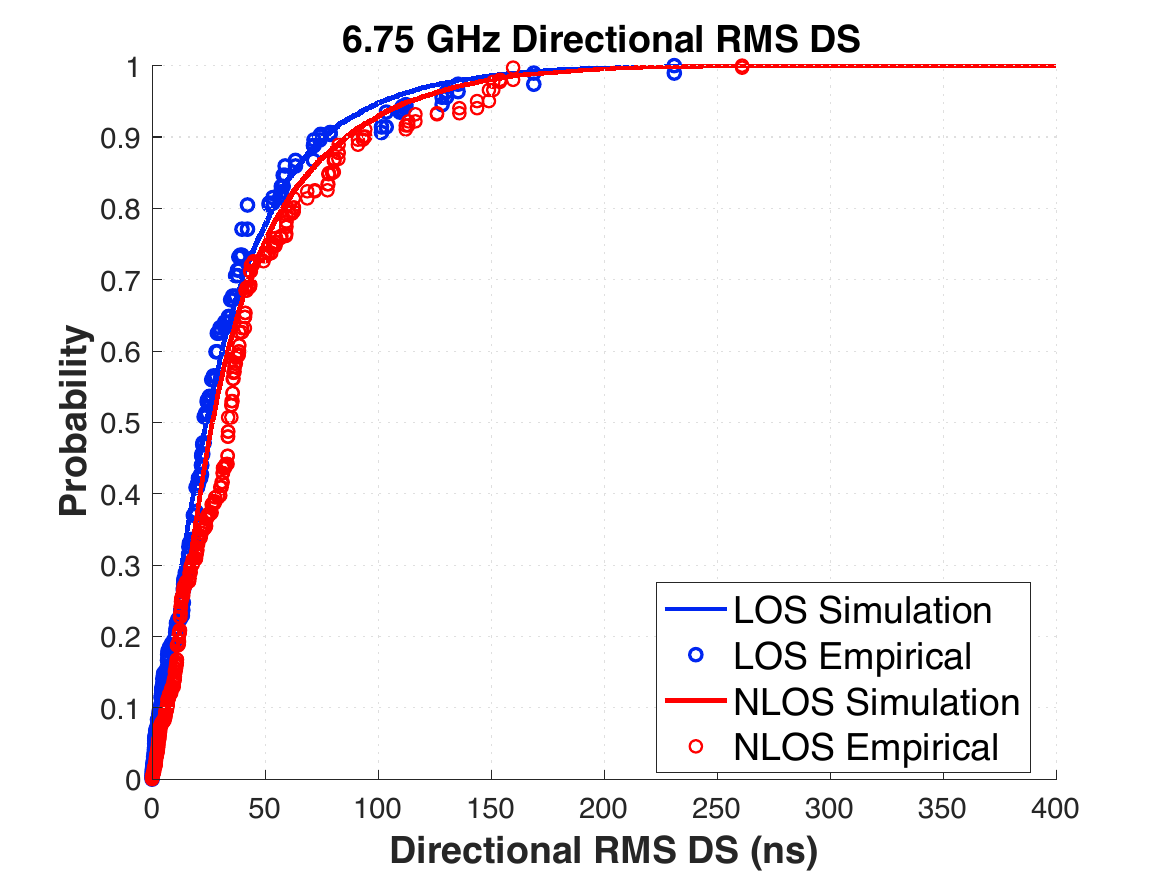}
      \vspace{-3pt}
  \caption{{CDF of directional RMS DS for 6.75 GHz. Measurements \cite{globec_indoors,icc_inf,globec_penetration} were conducted using a 1 GHz-bandwidth sliding-correlation channel sounder and directional horn antennas with 15 dBi gain and 30° HPBW.}}
  \label{fig:7G}
   \vspace{-5pt}
\end{figure}



 Figs~\ref{fig:17G} and \ref{fig:7G} illustrate examples of the measurement-based statistical channel model (SCM) developed in \cite{ojcoms_fr3_overview,globec_indoors,icc_inf,globec_penetration}, implemented in NYUSIM Python. In particular, Fig.~\ref{fig:17G} and Fig.~\ref{fig:7G} show CDFs for measured and simulated directional root mean square (RMS) delay spread (DS) for the 16.95 GHz and 6.75 GHz campaigns, respectively.
\begin{table}[!t]
\centering
\footnotesize
\setlength{\tabcolsep}{6pt}
\renewcommand{\arraystretch}{1}
\caption{Measured \cite{globec_indoors,icc_inf,globec_penetration} vs. Simulated Dir. RMS DS Mean and Std.}
\label{t1}
\begin{tabular}{lll*{4}{c}}
\toprule
\multicolumn{3}{c}{Dir. RMS DS} &
\multicolumn{2}{c}{\textbf{16.95~GHz}} &
\multicolumn{2}{c}{\textbf{6.75~GHz}} \\
\cmidrule(lr){4-5}\cmidrule(lr){6-7}
\multicolumn{3}{c}{$\log_{10}$[ns]} & Meas. & Sim. & Meas. & Sim. \\
\midrule
 & \multirow{2}{*}{$\scalebox{1.2}{$\mu$}_{\mathrm{RMS\,DS_{dir}}}$} & LOS  & 1.23 & 1.30 & 1.29 & 1.30 \\
 &                             & NLOS & 1.35 & 1.38 & 1.43 & 1.36 \\
 & \multirow{2}{*}{$\scalebox{1.2}{$\sigma$}_{\mathrm{RMS\,DS_{dir}}}$} & LOS  & 0.67 & 0.55 & 0.60 & 0.53 \\
 &                             & NLOS & 0.55 & 0.85 & 0.50 & 0.46 \\
\bottomrule
\end{tabular}
\vspace{-0.3cm}
\end{table}

The RMS DS is a measure of the temporal spread of multi-path components (MPCs)
in a channel. To evaluate directional DS, an arbitrary complex
antenna patterns $g_{\mathrm{TX}}$ and $g_{\mathrm{RX}}$ are applied to the omnidirectional channel impulse response (CIR) given by \cite[Eq.~(3)]{Dipankar_2024}, which results in the directional CIR to be of the following form \cite{Dipankar_2024}: 
\begin{IEEEeqnarray}{rCl}\label{eq:dir-cir}
h_{\mathrm{dir}}\!\left(t,\vec{\Theta},\vec{\Phi}\right)
&=& \sum_{n=1}^{N} \sum_{m=1}^{M_n}
a_{m,n}\, e^{j\varphi_{m,n}}\, \delta\!\left(t-\tau_{m,n}\right)\notag \\
&& g_{\mathrm{TX}}\!\left(\vec{\Theta}-\vec{\Theta}_{m,n}\right)\,
g_{\mathrm{RX}}\!\left(\vec{\Phi}-\vec{\Phi}_{m,n}\right),
\end{IEEEeqnarray}
where $t$ is the absolute propagation time, 
$\vec{\Theta} = \big(\phi_{\mathrm{AOD}},\,\theta_{\mathrm{ZOD}}\big)$
represents the TX pointing direction vector, and 
$\vec{\Phi} = \big(\phi_{\mathrm{AOA}},\,\theta_{\mathrm{ZOA}}\big)$
is the RX pointing direction vector. 
$N$ and $M_n$ denote the number of time clusters (TCs) and the number of spatial lobes (SLs), respectively; 
$a_{m,n}$ is the magnitude of the $m$th SL belonging to the $n$th TC, 
while $\varphi_{m,n}$ and $\tau_{m,n}$ represent the phase and propagation delay of the SL, respectively. 
Likewise, $\vec{\Theta}_{m,n}$ and $\vec{\Phi}_{m,n}$ are the vectors representing AOD/ZOD and AOA/ZOA for the SL, respectively.  In the simulation,  in each run, 10,000 directional channel impulse responses are generated, and the directional RMS DSs are then extracted. The statistical channel model follows the clustered delay model in \cite[Table VIII]{Dipankar_2024}, with exponential inter-cluster and intra-cluster delay distributions. We set $N_c = 4$, and for the 16.95 GHz,  $\mu_{s, LOS} = 30$, $\mu_{s,NLOS} = 32$ and for $6.75$ GHz, $\mu_{s, LOS} = 18$, and $\mu_{s,NLOS} = 22$.  Table~\ref{t1} shows the corresponding mean and standard deviation of the RMS DS in the base-$10$ logarithmic domain. Figs. \ref{fig:17G} and \ref{fig:7G} and Table~\ref{t1} show that  NYUSIM Python faithfully recreates the extensive field measurements reported in \cite{globec_indoors,icc_inf,globec_penetration}. Further work verifies that the Python code faithfully recreates the statistics of NYUSIM v4.0 MATLAB when running 10,000 runs to produce sample functions over many operating modes and environments.

\section{Conclusion and Future Work} 
This paper describes the evolution of NYUSIM as a wideband statistical, physically based, and measurement-driven channel simulator. The verified NYUSIM Python version preserves the accuracy and statistical behavior of the MATLAB implementation while establishing a reproducible foundation for large-scale simulation and integration with AI and ML for channel modeling. A standardized 3D antenna data format \emph{(Ant3D)} and the inclusion of FR3 propagation statistics extend the simulator’s real-world accuracy and faithful reproduction of realistic spatio-temporal channel impulse response models, enabling the study of spatial, angular, and polarization effects that dominate at FR3 and mmWave frequencies. 
Building on the existing integration with ns-3, future work will expand NYUSIM Python to model dynamic mobility, vehicular, and sub-THz channels, and to enable AI-based channel synthesis and parameter inference. Creating the new Python code base will strengthen the role of NYUSIM as a measurement-based and physically interpretable platform to advance 6G propagation, PHY-layer research, and system design.


\section*{Acknowledgment}

This work was supported by the NYU WIRELESS Industrial Affiliates Program, NSF Grant No. 2234123, and NYU Technology Acceleration and Commercialization Award. 

\bstctlcite{BSTcontrol}
\bibliographystyle{IEEEtran}
\vspace{-0.9mm}
\bibliography{referencesv02} 

@IEEEtranBSTCTL{BSTcontrol,
  CTLuse_forced_etal = "yes",
  CTLmax_names_forced_etal = 2,
  CTLnames_show_etal = 1,
  CTLdash_repeated_names = "no"
}

@techreport{3gpp38901,
  author       = {3GPP},
  title        = {Study on Channel Model for Frequencies from 0.5 to 100 {GHz}},
  institution  = {ETSI},
  number       = {TR 38.901 V17.0.0},
  year         = {2022}
}

@article{poddar2024tutorial,
  author       = {H. Poddar and S. Ju and D. Shakya and T. S. Rappaport},
  title        = {A Tutorial on {NYUSIM}: Sub-Terahertz and Millimeter-Wave Channel Simulator for {5G}, {6G}, and Beyond},
  journal      = {IEEE Commun. Surveys Tuts.},
  volume       = {26},
  number       = {2},
  pages        = {824--857},
  year         = {2024},
  doi          = {10.1109/COMST.2023.3344671}
}

@article{letaief2019roadmap,
  author       = {K. B. Letaief and W. Chen and Y. Shi and J. Zhang and Y. J. A. Zhang},
  title        = {The Roadmap to {6G}: {AI}-Empowered Wireless Networks},
  journal      = {IEEE Commun. Mag.},
  volume       = {57},
  number       = {8},
  pages        = {84--90},
  year         = {2019},
  doi          = {10.1109/MCOM.2019.1900271}
}

@article{COST2100,
  author       = {L. Liu and C. Oestges and J. Poutanen and K. Haneda and P. Vainikainen and F. Quitin and F. Tufvesson and P. De Doncker},
  title        = {The {COST 2100} {MIMO} Channel Model},
  journal      = {IEEE Wireless Commun.},
  volume       = {19},
  number       = {6},
  pages        = {92--99},
  year         = {2012}
}

@article{QuaDRiGa,
  author       = {S. Jaeckel and L. Raschkowski and K. B\"{o}rner and L. Thiele},
  title        = {{QuaDRiGa}: A 3-{D} Multi-Cell Channel Model with Time Evolution for Enabling Virtual Field Trials},
  journal      = {IEEE Trans. Antennas Propag.},
  volume       = {62},
  number       = {6},
  pages        = {3242--3256},
  year         = {2014}
}

@inproceedings{Samimi2015,
  author       = {M. K. Samimi and T. S. Rappaport},
  title        = {Statistical Channel Model with Multi-Frequency and Arbitrary Antenna Beamwidth for Millimeter-Wave Outdoor Communications},
  booktitle    = {Proc. IEEE GLOBECOM Workshops},
  pages        = {1--7},
  year         = {2015}
}

@article{Rappaport2015Com,
  author       = {T. S. Rappaport and G. R. MacCartney and M. K. Samimi and S. Sun},
  title        = {Wideband Millimeter-Wave Propagation Measurements and Channel Models for Future Wireless Communication System Design},
  journal      = {IEEE Trans. Commun.},
  volume       = {63},
  number       = {9},
  pages        = {3029--3056},
  year         = {2015}
}

@article{Rappaport2017TAP,
  author       = {T. S. Rappaport and Y. Xing and G. R. MacCartney and A. F. Molisch and E. Mellios and J. Zhang},
  title        = {Overview of Millimeter-Wave Communications for Fifth-Generation ({5G}) Wireless Networks---With a Focus on Propagation Models},
  journal      = {IEEE Trans. Antennas Propag.},
  volume       = {65},
  number       = {12},
  pages        = {6213--6230},
  year         = {2017}
}

@inproceedings{PoddarICC2023,
  author    = {H. Poddar and T. Yoshimura and M. Pagin and T. S. Rappaport and A. Ishii and M. Zorzi},
  title     = {Full-Stack End-to-End {mmWave} Simulations Using {3GPP} and {NYUSIM} Channel Model in {ns-3}},
  booktitle = {Proc. IEEE ICC},
  year      = {2023},
  month     = May,
  pages     = {1048--1053},
}

@inproceedings{PoddarWS2023,
  title={ns-3 implementation of sub-Terahertz and Millimeter wave drop-based NYU channel model {(NYUSIM)}},
  author={Poddar, Hitesh and Yoshimura, Tomoki and Pagin, Matteo and Rappaport, Theodore and Ishii, Art and Zorzi, Michele},
  booktitle={Proceedings of the 2023 Workshop on ns-3},
  pages={19--27},
  year={2023}
}

@article{Sulyman2014,
  author       = {A. Sulyman and A. Nassar and M. Samimi and G. MacCartney and T. Rappaport and A. Alsanie},
  title        = {Radio Propagation Path Loss Models for {5G} Cellular Networks in the {28~GHz} and {38~GHz} Bands},
  journal      = {IEEE Commun. Mag.},
  volume       = {52},
  number       = {9},
  pages        = {78--86},
  year         = {2014}
}

@techreport{NIST2019,
  author       = {D. Novotny and K. Remley and W. Young},
  title        = {Millimeter-Wave Channel Sounding for Over-the-Air Testing},
  institution  = {Nat. Inst. Standards Technol. (NIST)},
  number       = {Technical Note 2068},
  year         = {2019}
}

@inproceedings{Xing140GHz2021,
  author    = {Y. Xing and O. Kanhere and S. Ju and T. S. Rappaport},
  booktitle = {Proc. IEEE GLOBECOM},
  title     = {Indoor Wireless Channel Properties at Millimeter Wave and Sub-Terahertz Frequencies},
  year      = {2019},
  pages     = {1--6},
}

@inproceedings{xing2021outdoor,
  author    = {Y. Xing and T. S. Rappaport},
  title     = {Propagation Measurements and Path Loss Models for sub-{THz} in Urban Microcells},
  booktitle = {Proc. IEEE ICC},
  year      = {2021},
  pages     = {1--6},
  doi       = {10.1109/ICC42927.2021.9500385},
}

@article{ojcoms_fr3_overview,
  author  = {D. Shakya and M. Ying and T. S. Rappaport and H. Poddar and
             P. Ma and Y. Wang and I. Al-Wazani},
  journal = {IEEE Open J. Commun. Soc.},
  title   = {Comprehensive {FR1(C)} and {FR3} Lower and Upper Mid-Band
             Propagation and Material Penetration Loss Measurements and
             Channel Models in Indoor Environment for {5G} and {6G}},
  year    = {2024},
  volume  = {5},
  pages   = {5192--5218},
  doi     = {10.1109/OJCOMS.2024.3431686}
}

@inproceedings{globec_indoors,
  title     = {Propagation Measurements and Channel Models in Indoor Environment at 6.75 {GHz} {FR1(C)} and 16.95 {GHz} {FR3}},
  author    = {D. Shakya and M. Ying and T. S. Rappaport and others},
  booktitle = {Proc. IEEE GLOBECOM},
  year      = {2024},
  pages     = {998--1003}
}

@inproceedings{icc_inf,
  title     = {Upper Mid-Band Channel Measurements and Characterization at 6.75 {GHz FR1(C)} and 16.95 {GHz} {FR3} in an Indoor Factory Scenario},
  author    = {M. Ying and D. Shakya and T. S. Rappaport and others},
  booktitle = {Proc. IEEE ICC},
  year      = {2025},
  pages     = {3303--3308}
}

@inproceedings{globec_penetration,
  author    = {D. Shakya and M. Ying and T. S. Rappaport and H. Poddar and P. Ma and Y. Wang and I. Al-Wazani},
  title     = {Wideband Penetration Loss through Building Materials and Partitions at 6.75 {GHz} in {FR1(C)} and 16.95 {GHz} in the {FR3} Upper Mid-Band Spectrum},
  booktitle = {Proc. IEEE GLOBECOM},
  year      = {2024},
  pages     = {1665--1670},
  doi       = {10.1109/GLOBECOM52923.2024.10901400},
}

@article{oyie2018_14_22GHz,
  author  = {N. O. Oyie and T. J. O. Afullo},
  title   = {Measurements and Analysis of Large-Scale Path Loss Model at 14 and 22 {GHz} in Indoor Corridor},
  journal = {IEEE Access},
  volume  = {6},
  pages   = {17205--17214},
  year    = {2018}
}

@article{kim2014_11GHz,
  author  = {M. Kim and Y. Konishi and Y. Chang and J.-I. Takada},
  title   = {Large-Scale Parameters and Double-Directional Characterization of Indoor Wideband Radio Multipath Channels at 11 {GHz}},
  journal = {IEEE Trans. Antennas Propag.},
  volume  = {62},
  number  = {1},
  pages   = {430--441},
  year    = {2014}
}

@article{Dipankar_2024,
  author  = {D. Shakya and S. Ju and O. Kanhere and H. Poddar and Y. Xing and T. S. Rappaport},
  journal = {IEEE Trans. Antennas Propag.},
  title   = {Radio Propagation Measurements and Statistical Channel Models for Outdoor Urban Microcells in Open Squares and Streets at 142, 73, and 28 {GHz}},
  year    = {2024},
  volume  = {72},
  number  = {4},
  pages   = {3580--3595},
  doi     = {10.1109/TAP.2024.3366581}
}

@inproceedings{10437154,
  author    = {U. Sengupta and C. Jao and A. Bernacchia and S. Vakili and D.-S. Shiu},
  booktitle = {Proc. IEEE GLOBECOM},
  title     = {Generative diffusion models for radio wireless channel modelling and sampling},
  year      = {2023},
  pages     = {4779--4784},
  doi       = {10.1109/GLOBECOM54140.2023.10437154}
}

@article{Dong2019DeepCNN_mmWaveMIMO,
  author  = {P. Dong and H. Zhang and G. Y. Li and I. S. Gaspar and N. NaderiAlizadeh},
  title   = {Deep {CNN}-Based Channel Estimation for mmWave Massive {MIMO} Systems},
  journal = {IEEE J. Sel. Topics Signal Process.},
  year    = {2019},
  volume  = {13},
  number  = {5},
  pages   = {989--1000},
  month   = sep,
  doi     = {10.1109/JSTSP.2019.2925975}
}

@inproceedings{Abadi2016TensorFlow,
  author    = {M. Abadi and P. Barham and J. Chen and Z. Chen and A. Davis and J. Dean and M. Devin and S. Ghemawat and G. Irving and M. Isard and M. Kudlur and J. Levenberg and R. Monga and S. Moore and D. G. Murray and B. Steiner and P. Tucker and V. Vasudevan and P. Warden and M. Wicke and Y. Yu and X. Zheng},
  title     = {TensorFlow: A System for Large-Scale Machine Learning},
  booktitle = {Proc. 12th USENIX Symp. OSDI},
  year      = {2016},
  pages     = {265--283}
}

@article{Wu2024CDDM_TWC,
  author    = {T. Wu and Z. Chen and D. He and L. Qian and Y. Xu and M. Tao and W. Zhang},
  title     = {{CDDM}: Channel Denoising Diffusion Models for Wireless Semantic Communications},
  journal   = {IEEE Trans. Wireless Commun.},
  year      = {2024},
  volume    = {23},
  number    = {9},
  pages     = {11168--11183},
  month     = sep,
  doi       = {10.1109/TWC.2024.3379244}
}

@techreport{nyusimv4manual,
  title       = {{NYUSIM} Version 4.0 User Manual},
  author      = {T. S. Rappaport and S. Sun and G. R. MacCartney and M. K. Samimi and Y. Xing},
  institution = {NYU WIRELESS, NYU Tandon School of Engineering},
  year        = {2022},
  url         = {https://wireless.engineering.nyu.edu/nyusim/}
}

@article{SamimiTMTT2016_3DStatModel,
  author  = {M. K. Samimi and T. S. Rappaport},
  title   = {3-{D} Millimeter-Wave Statistical Channel Model for {5G} Wireless System Design},
  journal = {IEEE Trans. Microw. Theory Techn.},
  volume  = {64},
  number  = {7},
  pages   = {2207--2225},
  month   = jul,
  year    = {2016}
}

@inproceedings{RappaportVTC2017Compare3GPPNYUSIM,
  author    = {T. S. Rappaport and S. Sun and M. Shafi},
  title     = {Investigation and Comparison of {3GPP} and {NYUSIM} Channel Models for {5G} Wireless Communications},
  booktitle = {Proc. IEEE VTC-Fall},
  year      = {2017},
  pages     = {1--5},
  doi       = {10.1109/VTCFall.2017.8287877},
}

@article{SunTVT2018PropagationModels,
  author  = {S. Sun and T. S. Rappaport and M. Shafi and P. Tang and J. Zhang and P. J. Smith},
  title   = {Propagation Models and Performance Evaluation for {5G} Millimeter-Wave Bands},
  journal = {IEEE Trans. Veh. Technol.},
  volume  = {67},
  number  = {9},
  pages   = {8422--8439},
  month   = sep,
  year    = {2018}
}

@inproceedings{SunICC2017NYUSIM,
  author    = {S. Sun and G. R. MacCartney Jr. and T. S. Rappaport},
  title     = {A novel millimeter-wave channel simulator and applications for {5G} wireless communications},
  booktitle = {Proc. IEEE ICC},
  year      = {2017},
  pages     = {1--7},
}

@techreport{ITUM2101,
  title        = {Modelling and simulation of {IMT} networks and systems for use in sharing and compatibility studies},
  author       = {{ITU-R}},
  number       = {M.2101-0},
  type         = {Recommendation},
  institution  = {International Telecommunication Union (ITU-R)},
  address      = {Geneva, Switzerland},
  month        = feb,
  year         = {2017},
  pages        = {36}
}

@article{bazzi2025uppermidband,
  author  = {A. Bazzi and M. Chafii and T. S. Rappaport},
  title   = {Upper Mid-Band Spectrum for {6G}: Vision, Opportunity, and Challenges},
  journal = {IEEE Commun. Mag.},
  year    = {2025},
  note    = {to appear}
}

@techreport{oneill:pcg2014,
  title        = {{PCG}: A Family of Simple Fast Space-Efficient Statistically Good Algorithms for Random Number Generation},
  author       = {Melissa E. O'Neill},
  institution  = {Harvey Mudd College},
  address      = {Claremont, CA},
  number       = {HMC-CS-2014-0905},
  year         = {2014},
  month        = sep,
  xurl         = {https://www.cs.hmc.edu/tr/hmc-cs-2014-0905.pdf}
}

@book{Tranter2004Simulation,
  author    = {W. H. Tranter and K. S. Shanmugan and T. S. Rappaport and K. L. Kosbar},
  title     = {Principles of Communication Systems Simulation with Wireless Applications},
  publisher = {Prentice Hall},
  address   = {Upper Saddle River, NJ, USA},
  year      = {2004},
  isbn      = {0-13-494790-8}
}

\end{document}